\documentclass[3p,twocolumn,times]{elsarticle}
\journal{Future Generation Computer Systems}

\usepackage[subtle,tracking=normal]{savetrees} %
\usepackage{amsmath,amssymb,amsfonts}
\usepackage{algorithmic}
\usepackage{graphicx}
\usepackage{textcomp}
\usepackage[figuresright]{rotating}
\usepackage[table,dvipsnames]{xcolor}
\definecolor{Gray}{gray}{0.9}
\usepackage{colortbl}
\usepackage{multirow}
\usepackage{multicol}
\usepackage{array}
\usepackage{booktabs}
\newcommand{\midsepremove}{\aboverulesep=0mm \belowrulesep=0mm}
\midsepremove
\newcommand{\midsepdefault}{\aboverulesep=0mm \belowrulesep=0mm}
\midsepdefault
\usepackage{hyperref}
\usepackage[inline]{enumitem}
	\setlist[enumerate]{leftmargin=1em}%
	\setlist[itemize]{leftmargin=1em,label={$\bullet$}}%

\colorlet{myHighlight}{White!80!Green}
\colorlet{newContent}{White!80!Blue}
\usepackage{amsthm}
\usepackage{thmtools}
\declaretheorem[
    name=Finding,
    shaded={bgcolor=myHighlight,padding=3pt,rulecolor=ForestGreen,rulewidth=0.5pt},
]{finding}

\begin{document}

\begin{frontmatter}

\title{A Systematic Evaluation of the Potential of\\Carbon-Aware Execution for Scientific Workflows}

\author[uofg]{Kathleen West}\ead{kathleen.west@glasgow.ac.uk}
\author[uofg]{Youssef Moawad}\ead{youssef.moawad@glasgow.ac.uk}
\author[hu]{Fabian Lehmann}\ead{fabian.lehmann@hu-berlin.de}
\author[hu]{Vasilis Bountris}\ead{vasilis.bountris@hu-berlin.de}
\author[hu]{Ulf Leser}\ead{ulf.leser@hu-berlin.de}
\author[uofg]{Yehia Elkhatib}\ead{yehia.elkhatib@glasgow.ac.uk}
\author[uofg]{Lauritz Thamsen}\ead{lauritz.thamsen@glasgow.ac.uk}
\address[uofg]{University of Glasgow, United Kingdom}
\address[hu]{Humboldt-Universität zu Berlin, Germany}

\begin{abstract}
Scientific workflows are critical to scientific data analysis and often involve computationally intensive processing of large datasets on compute clusters. As such, their execution tends to be long-running and resource-intensive, resulting in significant energy consumption and carbon emissions. While carbon-aware computing methods have received considerable attention in general cloud contexts, their application to scientific data analysis workflows remains a critical research gap. Our study addresses this oversight by showing how the delay tolerance, interruptibility, and scalability of scientific workflows can be leveraged for a significantly more sustainable execution model.

In this study, we first quantify the problem of carbon emissions associated with running scientific workflows, and then demonstrate the transformative potential for carbon-aware workflow execution. We estimate the carbon footprint of seven real-world Nextflow workflows executed on diverse dedicated cluster and public cloud resources using high-resolution average and marginal grid carbon intensity data from open and commercial data providers. Furthermore, we conduct a systematic evaluation of the impact of carbon-aware temporal shifting, and the dynamic pausing and resuming of the workflow. Moreover, we investigate the impact of resource scaling at both workflow and workflow task levels. Finally, we report substantial potential reductions in overall carbon emissions, with temporal shifting capable of decreasing emissions by over 80\%, and resource scaling by 67\%. 
\end{abstract}


\end{frontmatter}

\section{Introduction}\label{sec:intro}
Scientists across domains rely on increasingly large datasets and complex workflows to perform, for example, image processing~\cite{berrimanMontageGridEnabled2004}, genome analysis~\cite{fellowsyatesReproduciblePortableEfficient2021}, and material simulations~\cite{schaarschmidtWorkflowEngineeringMaterials2021}. These scientific workflows are composed of orchestrated computational tasks~\cite{ludascherScientificWorkflows2009a}. Scientific workflow management systems (SWMS) such as Nextflow~\cite{ditommasoNextflowEnablesReproducible2017a} allow for the execution and monitoring of scientific workflows on distributed cluster infrastructure. 

Scientific workflows often process vast quantities of data in parallel across numerous cluster nodes, and thus tend to be resource-intensive with runtimes spanning hours to weeks~\cite{diasDatacentricIterationDynamic2015}. This leads to significant energy consumption and carbon emissions.
For example, the Galactic Plane project~\cite{deelmanPegasusCloudScience2016} ran 16 workflows that consumed 318,000 core hours to generate image mosaics. 
Similarly, an Earth observation workflow~\cite{lehmann2021force} showed runtime variations ranging from five to 81 hours per execution, depending on available resources, highlighting the need to assess and optimize the carbon footprint of such workflows. 

Prior initiatives to enhance the sustainability of scientific workflows have focused on improving energy efficiency through techniques such as energy-efficient scheduling~\cite{choudharyEnergyawareScientificWorkflow2022a, 9253551, 10098783, 5644899, DURILLO2014221}, and Dynamic Voltage and Frequency Scaling (DVFS)~\cite{choudharyEnergyawareScientificWorkflow2022a, cotes2017dynamic, safari2018energy, 9815507}. 
While these approaches can decrease energy consumption, they face increasing challenges from hardware constraints, with recent estimations suggesting that the increase in energy efficiency of devices has slowed to doubling only every 2.29 years~\cite{prieto2025evolution}.  
Techniques like DVFS are limited by device capabilities, i.e., it affects all tasks sharing CPU resources, requires privileged access, and is commonly unavailable in cloud environments. Moreover, the applicability of DVFS is limited due to lower limits for safe processor frequencies. 

More recent works aim to align computational loads with the availability of low-carbon energy through carbon-aware computing~\cite{wiesnerLetWaitAwhile2021c, 10.1145/3575693.3575709, 10.1145/3626788, 9770383, 10.1145/3620678.3624644}. This alignment can be achieved by temporally shifting and scaling flexible compute workloads against energy signals like carbon intensity (CI), which is a measure of the emissions produced per kilowatt-hour ($kWh$) of electricity consumed.
Temporal shifting involves scheduling applications to consume electricity when the CI is relatively low and to pause the workload otherwise~\cite{wiesnerLetWaitAwhile2021c, 9770383}. Resource scaling entails dynamically allocating resources to workloads based on the CI of electricity to make use of more resources when the CI is low, and to reduce demand when it is higher~\cite{10.1145/3626788, 10.1145/3575693.3575709}. 
There are two practically relevant CI signals: average and marginal. Average CI reflects the overall grid emissions, factoring in each energy source's relative share and emission rate. In contrast, marginal CI measures the emissions of the specific energy source meeting an additional load.
In many regions, both CI signals vary significantly due to intermittent renewables and demand fluctuations~\cite{wiesnerLetWaitAwhile2021c,he2021using}.

While these methods demonstrate the potential of carbon-aware computing, no study to date has systematically explored its application to scientific workflows across diverse applications, regions, and CI signals, using both carbon-aware scheduling and scaling methods. 
At the same time, scientific workflows appear particularly well-suited to carbon-aware computing owing to the following properties:
\begin{itemize}
    \item \emph{Delay tolerance}: Many scientific workflows will not have strict deadlines (e.g., executing against a new dataset or with a new algorithm when it becomes available). This allows time-shifting of executions based on low-carbon energy availability.
    \item \emph{Interruptibility}: Workflows are modeled as directed acyclic graphs of computational tasks that typically exchange intermediate results between tasks using disks, allowing to pause execution temporarily and to execute subsequent tasks from persisted data when lower carbon energy becomes available again.
    \item \emph{Scalability}: Resource allocation can be adjusted so that individual tasks are executed on machines of varying scales and entire workflows are run on clusters of different sizes. Furthermore, tasks can be embarrassingly parallel, allowing for parallel execution. 
    This enables the shaping of runtimes and resource usage against upcoming periods of low-carbon energy.
    \item \emph{Heterogeneity}: The tasks of a workflow can have varying resource demands, including possibly both CPU-intensive and I/O-intensive analysis steps, so energy-intensive tasks could utilize the lowest carbon energy available.
\end{itemize}

Addressing the identified gap, this paper rigorously and systematically assesses the potential of carbon-aware execution for scientific workflows. When evaluating the potential reduction in carbon emissions, we assume perfect knowledge of task and workflow executions, CI forecasts, and infinite resource availability. These assumptions serve to establish an upper bound on possible carbon savings when applying carbon-aware computing techniques. 

We expand significantly on preliminary results first presented in a short paper~\cite{11044837}, and we evaluate the potential emission savings for seven real-world workflows, implemented in Nextflow, and stemming from bioinformatics, remote sensing, and astronomy applications.
We assess carbon-aware temporal shifting of workflows, both with and without interruptions, alongside resource scaling at node and cluster levels.
Crucially, we quantify emissions and achievable savings by applying both average and marginal CI using real commercial-grade data. 
Through this comprehensive analysis, we make the following contributions:
\begin{itemize}
    \item We demonstrate the scale of the problem by estimating the carbon footprint of seven popular real-world Nextflow workflows from diverse scientific fields on varied cluster infrastructures.
    \item We systematically evaluate the potential of both carbon-aware temporal shifting and resource scaling of scientific workflows using average and marginal carbon intensity data.
    \item We provide our simulation and results analysis code as open source to enable reproducibility and future research~\footnote{\label{repo-footnote}\url{https://github.com/GlasgowC3lab/evaluate-carbon-aware-workflows}}.
\end{itemize}

\section{Background}
\label{sec:background}
We explain scientific workflows and carbon intensity.

\subsection{Scientific Workflows}
\begin{figure}[h]
\centerline{\includegraphics[width=0.8\columnwidth]{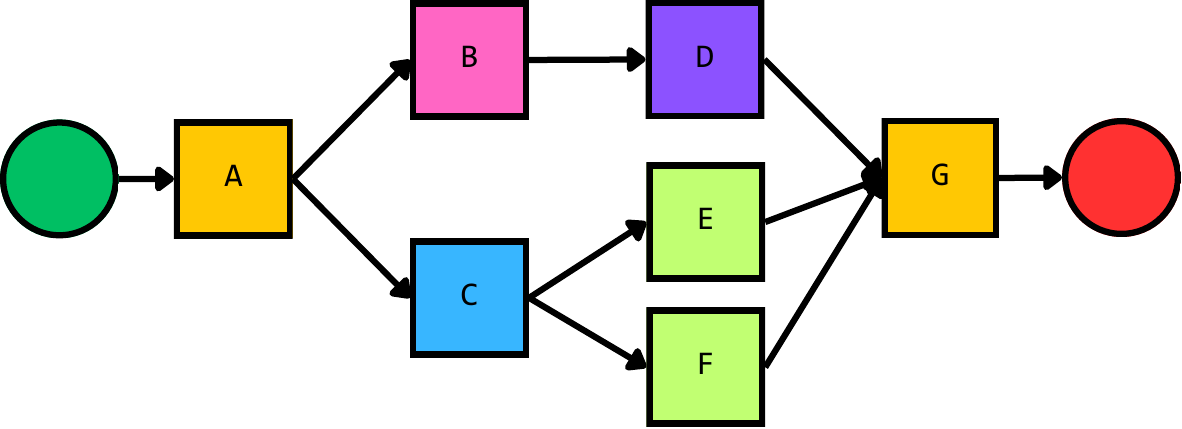}}
\caption{A scientific workflow formed of seven tasks.}
\label{fig:simple-workflow}
\end{figure}

Scientific workflows are typically depicted as directed acyclic graphs (DAGs). In these graphs, nodes represent computational tasks, and edges illustrate the data or control dependencies between them.
Scientific workflow systems automate the execution of workflows. 
Fig.~\ref{fig:simple-workflow} shows an example workflow, consisting of seven tasks that depend on each other; e.g., Task G requires input from Tasks D, E, and F.

Such tasks are typically self-contained programs that are often shared in binary form or as containers. Individual tasks are considered atomic and executed independently on a single machine. To decrease runtime, tasks can be executed on faster machines, and workflows can be executed on clusters in which more resources can be assigned. This enables some tasks to run in parallel when no dependency exists, such as Tasks B and C in Fig.~\ref{fig:simple-workflow}. 
In addition, workflows are often executed on multiple inputs, which allows for data-parallel execution of entire workflows on the allocated cluster resources.
Typically, the intermediate results that become inputs of subsequent tasks are exchanged through network file systems, providing flexibility to schedule subsequent tasks on different nodes as well as to recover from persisted intermediate results in case of task failures.

\subsection{Carbon Intensity (CI)}

\begin{figure*}[tb]
\centerline{\includegraphics[width=\textwidth]{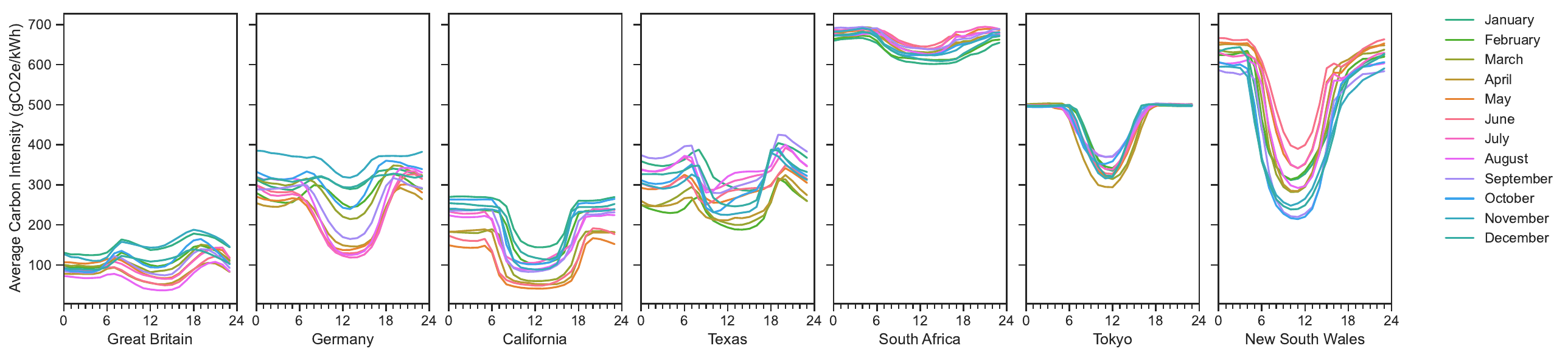}}
\caption{Daily mean average carbon intensity per month in 2024 for the seven regions we studied: Great Britain, Germany, California (USA), Texas (USA), South Africa, Tokyo (Japan), and New South Wales (Australia).}
\label{fig:carbon-intensity-fluctuation}
\end{figure*} 

CI measures the carbon emissions produced per unit of electricity consumed. Renewables have lower CI than fossil fuels, but their output levels vary over time. 
We analyze this fluctuation in seven regions during 2023 in Fig.~\ref{fig:carbon-intensity-fluctuation}, expanding on a previous analysis in the literature~\cite{wiesnerLetWaitAwhile2021c}. 

As CI can be quantified by the average or marginal signal, there has been an ongoing discourse on which signal should be used for carbon-aware optimizations~\cite{10.1145/3632775.3661953}. Average reflects the overall grid emissions, factoring in each energy source's relative share and emission rate. In contrast, marginal measures the emissions of the specific energy source meeting an additional load.
Marginal CI is preferred for measuring the impact of load shifting~\cite{he2021using}. However, there are challenges in obtaining the metric owing to computational complexity -- marginal CI is only estimated and lacks granularity for accurate reporting. 
In contrast, average CI can be measured and is often readily available and, therefore, commonly used for reporting. It could also help incentivize investment in renewable energy generation by aligning electricity usage with renewable sources for greater long-term impact~\cite{ryan2016comparative}.
As both signals have advantages and disadvantages, we consider both signals in our exploration of the potential impact of carbon-aware computing methods. 

\begin{figure}[tb]
\centerline{\includegraphics[width=0.9\columnwidth,trim={0 0.25cm 0 0.20cm},clip]{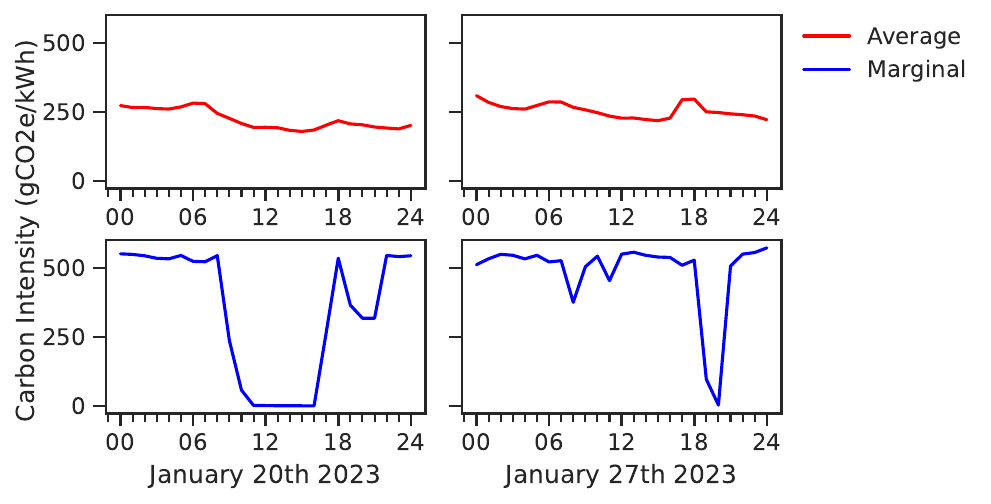}}
\caption{A comparison of average and marginal CI signals for Northern Texas between two days, 20th and 27th of January 2023.}
\label{fig:tx-ci-avg-vs-marg}
\end{figure}

Further, the marginal signal is highly variable and does not follow a predictable pattern in the same way that the average signal often does. 
In Fig.~\ref{fig:tx-ci-avg-vs-marg}, we plot the average and marginal CI on the 20\textsuperscript{th} and 27\textsuperscript{th} of January, 2023. On the same day a week apart, the average CI follows a similar pattern, whereas the marginal CI varies significantly. These dips could indicate windows of time in which electricity has a CI near zero, potentially due to renewable energy being curtailed. Curtailment is the deliberate reduction of electricity generation to balance supply and demand in a grid. It occurs when generation exceeds current grid demand, and the excess power cannot be stored or traded with neighbouring grids. 

\subsection{Operational vs. Embodied Carbon Emissions}
Operational carbon refers to emissions generated during the in-use phase of hardware, resulting from the electricity consumed to power computations, data storage, and networking. These emissions are ongoing and directly correlate with the system's energy efficiency, computational load, and the real-time carbon intensity of the electricity grid. 

In contrast, embodied carbon emissions are all emissions associated with the manufacturing and lifecycle of the physical hardware. This includes raw material extraction, component fabrication, transportation, assembly, and end-of-life activities such as e-waste processing and disposal. 
These stages are quantified by the Life Cycle Assessment (LCA) to measure the carbon footprint of computing devices~\cite{LCACritique1995}. 
While operational emissions can be mitigated dynamically through strategies like energy-efficient scheduling and carbon-aware load shifting, embodied emissions are a front-loaded carbon investment, set when the hardware is manufactured.

In this work, our focus is primarily on operational carbon emissions. However, as a first step towards a full picture of attributing emissions to computational workflows, we also include an estimate of the embodied emissions of utilizing the CPU while the workflows are executing. We demonstrate the effect of the different optimization techniques on the attributed embodied carbon emissions.

\section{Study Design}
\label{sec:study-design}
We capitalize on the fact that scientific workflows are particularly suited to carbon-aware computing, given their delay tolerance, interruptibility, scalability, and heterogeneity. 
In this paper, we focus on systematically exploring delay tolerance, interruptibility, and scalability leaving the exploitation of heterogeneity for future work. 
We focus on the \textbf{operational} emissions of carbon-aware temporal shifting and resource scaling, but also analyze the impact that optimizations have on \textbf{embodied} emissions.

For all experiments, we use existing methods and tools to estimate the energy consumption (based on resource utilization with linear power models), and then translate that into operational and embodied carbon emissions of workflows (based on commercial-grade average and marginal CI data). 
The novelty, therefore, lies in systematically evaluating the impact of previously proposed carbon-aware computing methods for specifically scientific workflows, which has not been done before.

To explore the potential reduction in carbon emissions, we make the following assumptions:
\begin{enumerate}[noitemsep]
    \item perfect knowledge of workflow task executions %
    \item CI forecasts are available without any error%
    \item infinite resource availability when performing time shifting and resource scaling
\end{enumerate}
The identified potential reduction can, consequently, differ from the actual reduction that can be achieved in practice. We explore the impact of uncertainty in workflow task runtimes and CI forecasts by performing sensitivity analyses in Section~\ref{subsec:SensitivityAnalyses} and further discuss our assumptions in Section~\ref{Sec:threats}.

We describe the experimental setup used in the following evaluation sections. This includes: (1) the selection of real-world Nextflow workflows, (2) the compute resources on which workflows and their tasks were executed, and (3) the CI data for the region where these were executed. 

\subsection{Scientific Workflows}
We study five of the ten most popular real-world bioinformatics workflows from Nextflow's community-curated nf-core library~\cite{ewels2020nf}, an astronomy workflow, and another from the Earth observation domain. These are representative of the domains they work with. Table~\ref{table:workflow-list} details these workflows. 

\begin{table}[htb]
\caption{The seven real-world workflows selected for investigation.}
\label{table:workflow-list}
\centering
\rowcolors{2}{}{Gray}
\resizebox{0.9\columnwidth}{!}{
\begin{tabular}{lcr}
\toprule
Workflow & Domain & \# Physical Tasks  \\
\midrule
Chip-Seq~\cite{harshil_patel_2024_13899404} & Bioinformatics & 3,536  \\
MAG~\cite{10.1093/nargab/lqac007} & Bioinformatics & 7,477 \\ 
Montage~\cite{berrimanMontageGridEnabled2004} & Astronomy & 197 \\  
Nano-Seq~\cite{harshil_patel_2023_7716033} & Bioinformatics & 91  \\
Rangeland~\cite{lehmann2021force} & Earth Observation & 4,417 \\   
RNA-Seq~\cite{harshil_patel_2024_13986791} & Bioinformatics & 1,268  \\ 
Sarek~\cite{10.1093/nargab/lqae031, 10.12688/f1000research.16665.2} & Bioinformatics  & 8,954 \\ 
\bottomrule
\end{tabular}
}
\end{table}

To minimize carbon emissions in our study, we rely on existing historical traces wherever possible. 
Specifically, we used trace files~\cite{wow_Results} for MAG and Rangeland~\cite{10678682} and for Chip-Seq, Montage, RNA-Seq, and Sarek from~\cite{lehmannWOW2025}. Meanwhile, we ran the Nano-Seq workflow on an edge server, as well as several individual workflow tasks on cloud and cluster resources. 

\subsection{Compute Resources}
In this study, we work with various nodes, which are detailed in Table~\ref{table:compute-resources}. 
We used three types of Google Compute Platform (GCP) nodes, all prefixed with ``gcp'', and five other types of nodes -- some of which are individual edge servers while others like -- olympus, atlantis and camelot are part of homogeneous clusters. 
These resources represent diverse and relevant compute environments for scientific workflow execution.

\begin{table}[htb]
\caption{The compute resources used in the study, and their associated LCA emissions.}
\label{table:compute-resources}
\centering
\rowcolors{3}{Gray}{}
\resizebox{\columnwidth}{!}{
\begin{tabular}{llrr}
\toprule
& & Memory & LCA Emissions\\
Name & Hardware & (GB) & ($kgCO2e$)               \\
\midrule
gcp-c2 & c2-standard-8 & 32  & 19.00 \\
gcp-n1 & n1-standard-2 & 7.5 & 19.00 \\
gcp-n2 & n2-highmem-8 & 32 & 19.00  \\ 
\hline
atlantis & AMD EPYC 7282 & 128 & 23.17 \\ 
camelot & Intel Xeon Silver 4314 & 256 & 21.00 \\
elysium & Intel Xeon Gold 6426Y & 128 & 46.73 \\ 
olympus & Intel Xeon E5-2640 & 64 & 19.80 \\
sherwood & Intel i7-10700T & 32 & 12.37 \\ 
\bottomrule
\end{tabular}
}
\end{table}

\subsection{Energy Consumption Estimation}
In our experiments, we estimate the carbon footprint from executing scientific workflows and their tasks. To achieve this, we used Ichnos~\cite{west2025ichnoscarbonfootprintestimator}. 
It is a tool we built to estimate the carbon footprint of Nextflow workflows from workflow traces, and allows users to provide power models for the compute resources utilized. We utilized Ichnos's implementation of a linear power model to estimate the energy consumption of resources, translating this to a carbon footprint with fine-grained CI data aligning with each workflow's execution. 
Ichnos enables post-hoc energy consumption estimation. When we compared the actual energy consumed, monitored using RAPL, we found that Ichnos' estimations were more accurate (4--10\% error) than other estimation methodologies like CCF~\footnote{\label{ccf-footnote}\url{https://www.cloudcarbonfootprint.org}} (14--48\% error) or GA~\footnote{\label{ga-footnote}\url{https://www.green-algorithms.org}} (81--98\%). 

\subsection{Carbon Intensity Data}
We performed all footprint estimations using average and marginal CI data sourced from Electricity Maps\footnote{\label{foot:ElectricityMaps}\url{https://www.electricitymaps.com/data-portal}} and WattTime\footnote{\label{foot:WattTime}\url{https://watttime.org/}}.
Given that Electricity Maps' average CI data was offered with hourly intervals, and WattTime's marginal CI data was offered with 5m intervals, we used the most granular CI data available for associated experiments, i.e., 5m for marginal CI and 60m for average CI. 

We selected seven regions, including those where the workflows were originally executed and others where electricity was generated from different renewable sources, prioritizing regions that have a significant data center presence. We selected regions from each continent to increase our representativeness. 
We selected Great Britain and Germany as they were the regions in which the scientific workflows were originally executed. We selected Texas, as it has led the US in energy generation from wind renewables; California, as it is the highest solar power generating state in the US; along with New South Wales, Tokyo and South Africa, as they, similar to Texas and California, have a significant data center presence as well as variable renewable energy sources. 

\subsection{Embodied Carbon Data and Attribution}
\label{sec:embodied_data}
We use the Boavizta API~\cite{Boavizta2025} to obtain LCA emissions of hardware components. This is an open-source toolkit that estimates embodied environmental impacts by modeling hardware components, manufacturing characteristics, and their allocation to cloud platforms using crowd-sourced and manufacturer-published LCA data. We attribute the embodied carbon for a component based on the runtime as a share of the device's expected lifetime -- 4 years for CPUs, following the CCF methodology~\textsuperscript{\ref{ccf-footnote}}).

As this is a first step towards studying the impact carbon-aware computing techniques have on the embodied carbon emissions of workflows, we only consider the impact of CPUs. However, we include a brief discussion of the impact of hard drive storage during interrupted workflow execution in Section~\ref{Subsec:Interruptibility}.

\section{Carbon Footprint Estimation for Workflows}
\label{sec:footprint-estimation}
To further motivate our paper's focus, as well as to establish baselines for the subsequent experiments, we first estimate the operational and embodied carbon emissions generated from the original workflow executions. 

We begin by estimating the carbon footprint for the Chip-Seq, MAG, Montage, Nano-Seq, Rangeland, RNA-Seq, and Sarek workflows. These estimations are collated in Table~\ref{table:workflows-analysis}. 
They are based on the mean of three executions of each identified workflow. We present each workflow's energy consumption on the utilized compute resources. We then translate this into carbon emissions, using either average and marginal CI data, based on the original start and execution times.

\begin{table*}[h]
\caption{Operational and embodied (Emb.) carbon emissions (emis) estimated for the selected workflows' original execution, using average (Avg.) and marginal (Marg.) CI.}
\label{table:workflows-analysis}
\centering
\rowcolors{3}{Gray}{}
\begin{tabular}{lcrrrr}
\toprule
& & Energy & Avg. emis & Marg. emis & Emb. emis \\
Workflow & Resources & (kWh) & (gCO2e) & (gCO2e) & (gCO2e) \\
\midrule
Chip-Seq & atlantis x8 & 7.70 & 4,243.75 & 5,958.90 & 117.89 \\
MAG     & camelot x8 & 21.12 & 4,301.94 & 15,015.20 & 213.10 \\
Montage & atlantis x8 & 0.46 & 244.07 & 342.89 & 5.05 \\
Nano-Seq & sherwood & 0.41 & 35.16 & 164.30 & 2.82      \\
Rangeland & camelot x8 & 11.46 & 2,876.44 & 8,280.52 & 70.23 \\
RNA-Seq  & atlantis x8 & 5.43 & 1,415.65 & 4,181.25 & 74.52 \\
Sarek   & atlantis x8 & 10.16 & 3,930.06 & 7,833.18 & 225.00 \\
\bottomrule
\end{tabular}
\end{table*}

Energy consumption varied significantly across workflows executed and resources utilized. However, the carbon emissions produced from each execution also depend on when and in which region workflows were executed. In this paragraph, we focus on estimating the produced carbon emissions using the average CI signal. 
Nano-Seq was executed in Great Britain, producing emissions of 85.8 $g/kWh$ of energy consumed. Montage was executed in Germany, producing 530.6 $g/kWh$. These rates depend on how carbon-intensive electricity in each region is, e.g., Great Britain being notably lower than Germany. 

If we compare two different workflows that were executed in Germany on the same compute resources, we see that their emission rate differs significantly. While RNA-Seq produced emissions of 260.7 $g/kWh$, Chip-Seq produced 551.1 $g/kWh$. These rates differ due to the CI fluctuating over time, with both workflows running at different times. 

These estimations account for the energy consumption of an entire workflow execution which encompasses: static energy consumed by the CPU and memory, dynamic energy consumed by each task (considering their runtime and CPU utilization), and energy consumed from allocating memory to each task.
We do not account for energy consumed when our workflow is not running, as the resources could be available to other workloads.

\footnotetext{\url{https://www.epa.gov/energy/greenhouse-gas-equivalencies-calculator}, Accessed December '25.}

Executing these workflows resulted in up to 4.3$\mathrm{kg}$ of carbon emissions for a single run (up to 13$\mathrm{kg}$ for three executions). For comparison, 4.3$\mathrm{kg}$ of carbon emissions is equivalent to the greenhouse gas emissions produced by driving 18$\mathrm{km}$ in an average petrol-powered car\footnotemark.

\section{Potential of Carbon-Aware Load Shifting}
\label{sec:load-shifting-potential}
In this section, we explore how entire workflow applications can be temporally shifted, how workflows can be paused and resumed to further reduce their carbon footprint, and the impact of temporal shifting in seven regions around the world.

\subsection{Entire Workflow Shifting} 
\label{Subsec:DelayTolerance}
In our first experiment, the start time of an entire workflow's execution is systematically adjusted by an hour, for every hour within a specified ``flexibility window'' to measure the potential reduction possible without further adjusting the workflow's execution. 

To ensure that our results were comparable and that we considered the changing seasons of the year, we shifted each workflow's start time to 9AM on the second Monday of each month in 2024. We then considered two such flexibility windows from this start time: one of 24 hours and the other of 96 hours. This is to mimic the scenario where scientists could delay starting their workflow for up to a day, or over the working week. 
We explored the possible reduction using average and marginal CI signals, with the full results online~\textsuperscript{\ref{repo-footnote}}. We performed the experiment for all seven regions, and discuss the impact of the entire workflow shifting in selected regions. 

\paragraph{Results Interpretation}
In the figures that follow, we present the maximum possible reduction in footprint of each workflow (on the y-axis), for each month of the year (on the x-axis) -- this reduction is shown for a 24h window on the left, and a 96h window on the right. The heatmap shows reductions according to the shade of green, with darker shades meaning greater reductions. The percentage reduction is also shown on the heatmap. A reduction of 100\% is denoted as ``X''. 

\begin{figure}[th]
\centerline{\includegraphics[width=\columnwidth,trim={0 0 2.5cm 0},clip]{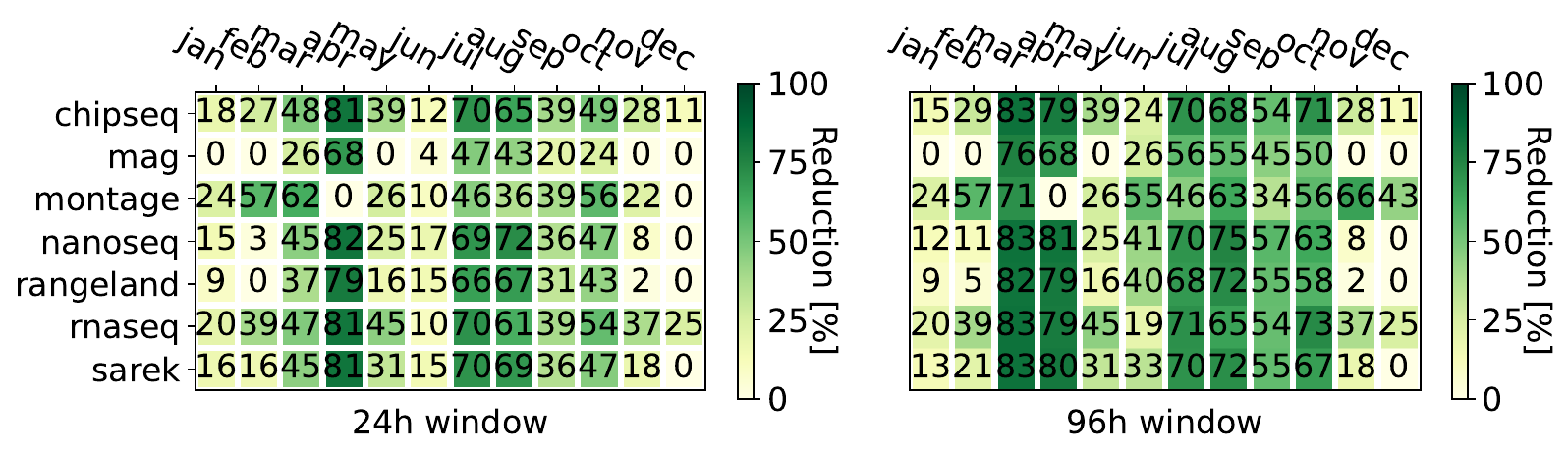}}
\caption{Reduction using entire workflow shifting in Great Britain}
\label{fig:entshift-gb-avg}
\end{figure}

\paragraph{Average CI}
In Fig.~\ref{fig:entshift-gb-avg}, we show the reduction possible using the average CI in Great Britain, which has a significant renewables presence. We see that longer shifting windows enable greater carbon reductions, with most workflows responding well to entire workflow shifting. 

\begin{figure}[th]
\centerline{\includegraphics[width=\columnwidth,trim={0 0 2.5cm 0},clip]{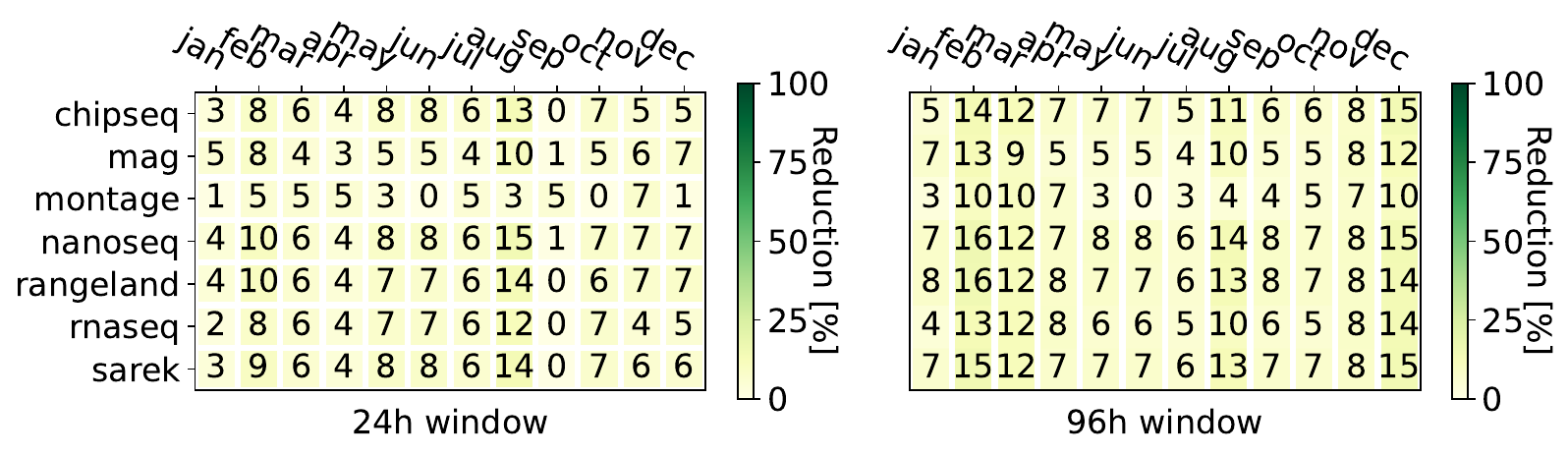}}
\caption{Reduction using entire workflow shifting in South Africa}
\label{fig:entshift-zaf-avg}
\end{figure}

However, the reduction potential depended on where and when the original workflows were executed, and the CI levels of the surrounding weekdays. For example, Fig.~\ref{fig:entshift-zaf-avg} shows the reduction potential in South Africa, using the average CI signal. Here, we see that there is little to no benefit from entire workflow shifting in either window. The region of South Africa's has a consistently high CI, with notably lower variability than other regions examined in this study (Fig.~\ref{fig:carbon-intensity-fluctuation}). 

\paragraph{Marginal CI}
When we performed the same experiments with the marginal CI signal, the potential impact of the entire workflow shifting was highlighted further. Given that the marginal signal is likely to indicate periods of time where the grid is curtailing renewable energy generation, or under low demand, we can observe a CI near zero over these periods. 

\begin{figure}[h]
\centerline{\includegraphics[width=\columnwidth,trim={0 0 2.5cm 0},clip]{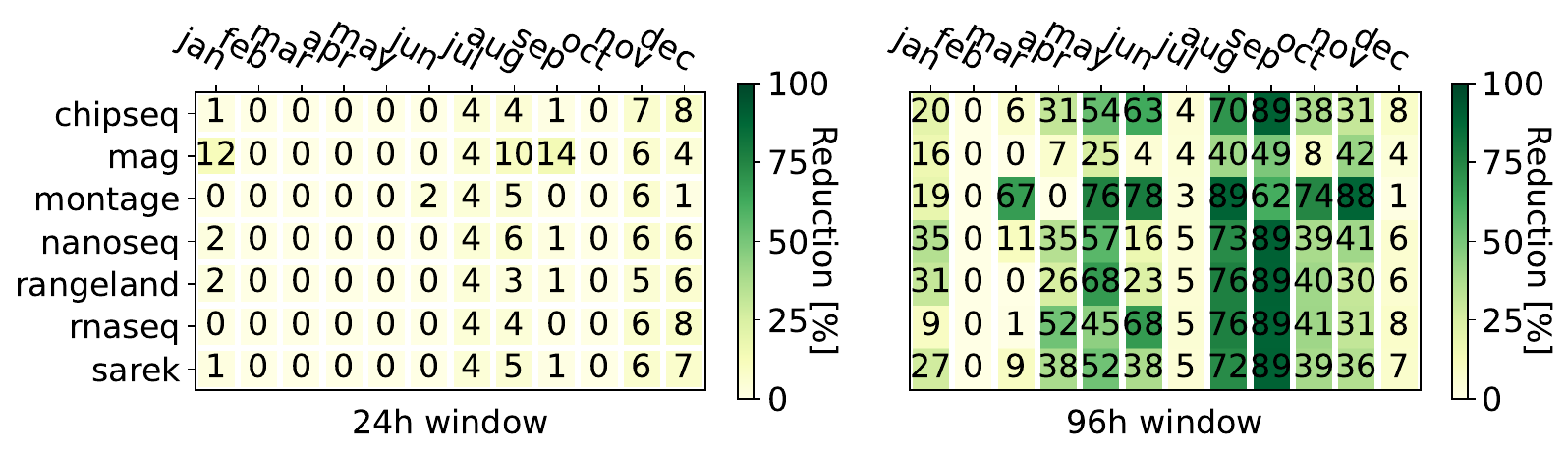}}
\caption{Reduction using Entire Workflow Shifting in California}
\label{fig:entshift-ca-marg}
\end{figure}

In Fig.~\ref{fig:entshift-ca-marg}, we show the reduction possible for California, which has a significant solar renewable generation presence. Here, we see that there is little-to-no reduction in a 24h window; but this increases significantly in a 96h window, most notably in September, where there was a period of low CI. This highlights the potential for reductions with increased flexibility of longer time horizons.

\begin{figure}[h]
\centerline{\includegraphics[width=\columnwidth,trim={0 0 2.75cm 0},clip]{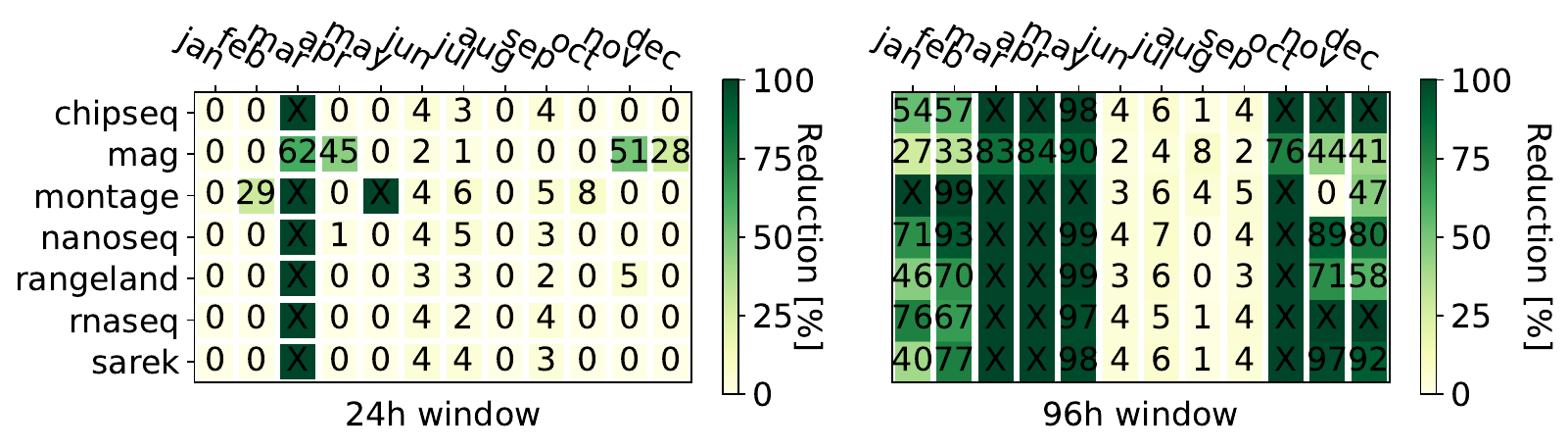}}
\caption{Reduction using entire workflow shifting in Texas}
\label{fig:entshift-tx-marg}
\end{figure}

In Fig.~\ref{fig:entshift-tx-marg}, we show the reduction possible for Texas, which has a significant wind renewable generation presence. We similarly observe that increasing the length of the window offers significant benefits for all workflows in several months of the year. In particular, we see that in some months like May, we can reduce the footprint by more than 90\%. However, this is reliant on the renewable signal capturing appropriate periods of low-carbon energy, and those periods occurring in a given region. 

\paragraph{Summary}
In this experiment, we observed that the potential reduction in emissions depends on where and when workflows could be executed, as the CI fluctuation is more pronounced in some regions with a greater renewables presence, while others may offer little benefit. Increasing the duration of the flexibility window tends to yield further reductions especially in regions with greater CI fluctuations. 

\paragraph{Impact on Embodied Emissions}
When we perform entire workflow shifting, we delay a workflow's execution, but do not alter its runtime or resource allocation. Consequently, we observe \textbf{no changes} to the embodied carbon footprint when compared to the baseline executions shown in Table~\ref{table:workflows-analysis}.

\subsection{Sensitivity Analysis}
\label{subsec:SensitivityAnalyses}
As explained in Section~\ref{sec:study-design}, we assumed access to CI forecasts with no error and perfect knowledge of workflow task executions. In practice, CI can be forecasted with a 3--20\% error~\cite{10.1145/3607114.3607117}, and workflow task runtimes can be predicted with an error of 3--30\%~\cite{BADER2024171, hilmanTaskRuntimePrediction2018}. 
We conducted sensitivity analyses modeling these realistic forecasting and prediction errors considering our entire workflow shifting experiment. Both analyses used average CI data for Great Britain. We report the emissions reduction deviation, which is the shortfall between the emissions reduction achievable with perfect data and the reduction possible when scheduling decisions rely on CI forecasts or task runtime predictions with an error. %

\paragraph{Carbon Intensity Forecasting}
We used CI data with a forecast error of 5, 10, and 15\%. Table~\ref{table:ci-sensitivity-analysis} shows the results.
When the CI forecast had an error, we saw a 2\% reduction deviation in emissions compared to the optimal reduction using actual data in the 24h window, and a 4.3\% reduction deviation in the 96h window. While the optimization is sensitive to forecasting errors, the reduction possible remains significant.

\begin{table}[htb]
\caption{Footprint Reduction Deviation when CI forecast has an error in 24h and 96h shifting windows.}
\label{table:ci-sensitivity-analysis}
\centering
\rowcolors{2}{}{Gray}
\resizebox{\columnwidth}{!}{
\begin{tabular}{l|ccc|ccc}
\toprule
& \multicolumn{3}{c|}{24h} & \multicolumn{3}{c}{96h} \\
CI Forecast Error (\%) & 5 & 10 & 15 & 5 & 10 & 15 \\
\midrule
Emissions Reduction Deviation & 0.3 & 0.9 & 2.0 & 0.5 & 1.9 & 4.3 \\ %
\bottomrule
\end{tabular}
}
\end{table}

\paragraph{Workflow Task Runtime Prediction}
We used workflow traces where task runtimes had a prediction error of 10, 15\%, and 25\%. The results are shown in Table~\ref{table:runtime-sensitivity-analysis}. 
We observed that there was minimal change compared to the optimal reduction in emissions (34\% and 43.5\%) as a result of workflow task runtime prediction errors.

\begin{table}[htb]
\caption{Footprint Reduction Deviation when task's predicted runtime has an error in 24h and 96h shifting windows.}
\label{table:runtime-sensitivity-analysis}
\centering
\rowcolors{2}{}{Gray}
\resizebox{\columnwidth}{!}{
\begin{tabular}{l|ccc|ccc}
\toprule
& \multicolumn{3}{c|}{24h} & \multicolumn{3}{c}{96h} \\
Runtime Prediction Error (\%) & 10 & 15 & 25 & 10 & 15 & 25 \\
\midrule
Emissions Reduction Deviation & 0.03 & 0.04 & 0.03 & 0.00 & 0.01 & 0.01 \\ %
\bottomrule
\end{tabular}
}
\end{table}

\subsection{Interrupted Workflow Shifting} 
\label{Subsec:Interruptibility}
In our second experiment, we considered how scientific workflows could be interrupted to exploit multiple shorter periods of low-carbon energy. 
For this experiment, we reflect that individual tasks cannot generally be paused and resumed, but that their start can be delayed without significant overhead. As workflow systems like Nextflow use disk storage to exchange intermediate results, there will be negligible runtime overhead for reading the inputs of tasks from disks at a later point in time.
The overhead of pausing and resuming entire workflow applications, hence, mainly stems from having to align task executions with multiple shorter periods of low-carbon energy availability, so that all tasks executed in a given time period finish fully within the given periods.

\begin{figure}[h]
\centerline{\includegraphics[width=0.48\columnwidth,trim={0 0.28cm 0 0},clip]{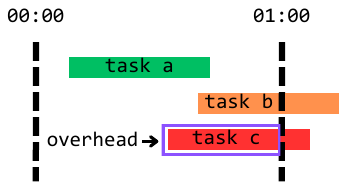}}
\caption{Defining hourly execution windows for workflow tasks and overheads.}
\label{fig:interrupt-overhead-eg}
\end{figure}

\paragraph{Overhead Estimation} 
As we used hourly granularity for CI data from both Electricity Maps and WattTime, we divided tasks from each entire workflow's execution into hourly windows as illustrated in Fig.~\ref{fig:interrupt-overhead-eg}. These windows contain two types of tasks:
\begin{enumerate*}[label=(\roman*)]
\item complete tasks that start and finish in the current hour, e.g., task a; and 
\item partial tasks that start in the current hour but finish later, e.g., tasks b and c. 
\end{enumerate*} 

Dividing tasks into hourly windows allows their execution to be aligned with multiple non-consecutive low-carbon windows of the CI time series. 
However, interrupting workflows introduces overhead, as tasks unable to finish within an hourly window have to be delayed to a later window. As multiple tasks are possibly delayed in this way, we can consider the task that is most delayed, that is, the longest partial task that runs within the window (the purple box around task c), as an upper bound of overhead. The overall overhead is the sum of the overheads of individual windows for every interval where an interruption occurred. 

We mapped the task execution windows to the lowest carbon intervals in a given flexibility window, in chronological order, to align with the workflow's original execution and data dependencies. 
Our results for average and marginal CI explore the potential reduction in carbon emissions for our selected workflows, highlighting the potential for temporal shifting in each original execution environment. In each of our selected regions, we explored the reduction possible by applying temporal shifting with interruptions. The full results are available online~\textsuperscript{\ref{repo-footnote}}. 

\begin{figure}[h]
\centerline{\includegraphics[width=\columnwidth]{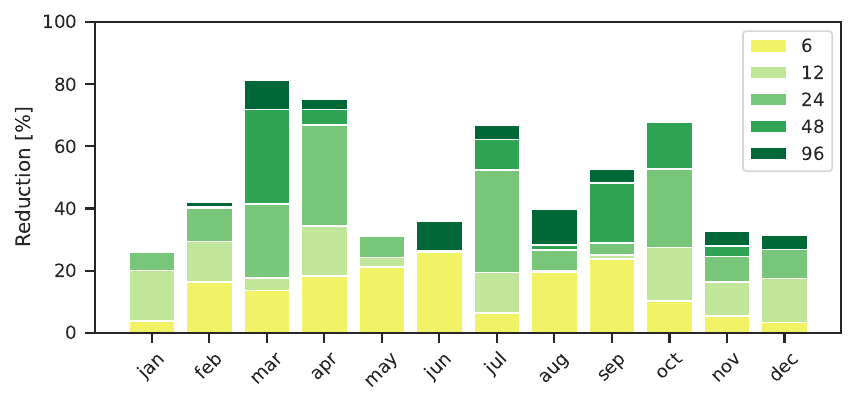}}
\caption{Footprint Reduction using Interrupted Shifting in Great Britain over windows of 6--96h.}
\label{fig:intshift-gb-avg}
\end{figure}

\paragraph{Results Interpretation}
In each graph, we show the percentage reduction in the carbon footprint, taking the mean reduction for our selected workflows with bar plots (on the y-axis). We calculated this reduction for each month of the year, using the same week as in the entire workflow shifting experiment (Section~\ref{Subsec:DelayTolerance}). Each bar plot is formed of five blocks, representing the reduction possible in each of five flexibility windows. For example, in Fig.~\ref{fig:intshift-gb-avg}, we see that in January, we could save around 20\% with the 12-hour window. Increasing the window to 24h, savings are increased to 25\%. Increasing the window size further does not yield additional reductions.   

\paragraph{Average CI}
In Fig.~\ref{fig:intshift-gb-avg}, which shows the reduction possible using interrupted shifting in Great Britain, we see that we can reduce the footprint by over 20\% in each month of the year, with much greater reductions possible by extending the flexibility window to at least 48h in some months of the year. We also see that the additional benefit in waiting 96h is small, and is only found in some months of the year. While these figures are similar to those from the entire workflow shifting experiment, we can achieve similar savings in a window of just 48h instead of 96h. 

\begin{figure}[h]
\centerline{\includegraphics[width=\columnwidth]{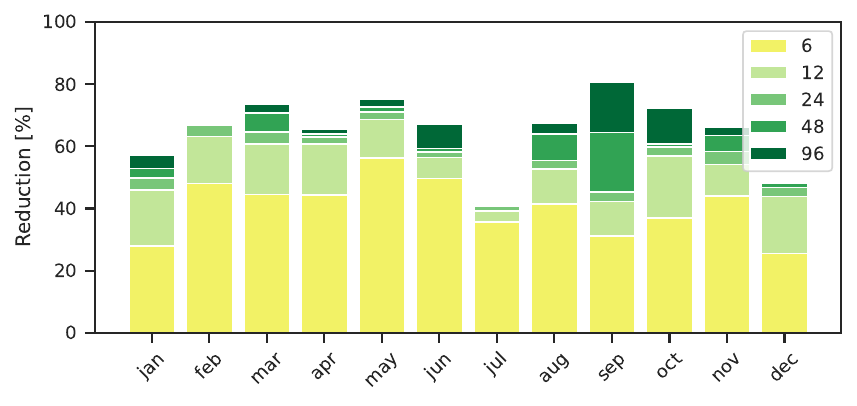}}
\caption{Footprint Reduction using Interrupted Shifting in California over windows of 6--96h.}
\label{fig:intshift-ca-avg}
\end{figure}

In Fig.~\ref{fig:intshift-ca-avg}, we show the reduction possible in California. Here, we see that our workflows footprint can be reduced by 40--65\% across all months of the year in a flexibility window of just 6--12h. Such underlines the potential of interrupted shifting over a short window, such as waiting from the morning until the evening, or from night to day. 
In contrast, using entire workflow shifting in the same region offers far lower reduction potential -- showing the benefit of interruptions in a region with a significantly variable solar renewable generation. 

\begin{figure}[h]
\centerline{\includegraphics[width=\columnwidth]{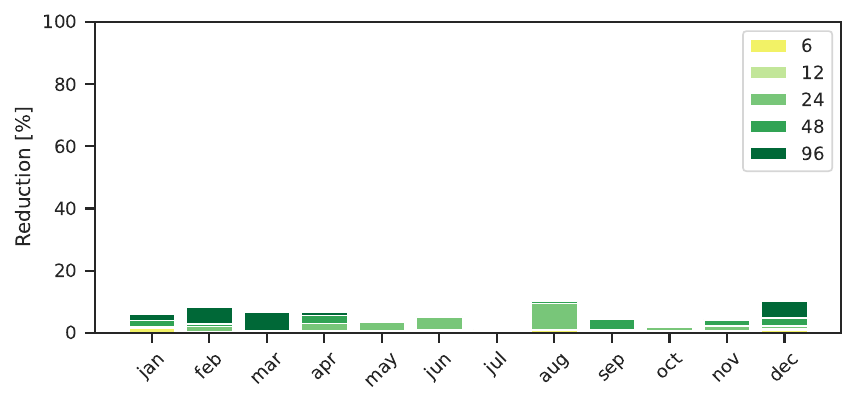}}
\caption{Footprint Reduction using Interrupted Shifting in South Africa over windows of 6--96h.}
\label{fig:intshift-zaf-avg}
\end{figure}

However, some regions, such as South Africa (see Fig.~\ref{fig:intshift-zaf-avg}) have relatively steady CI due to being heavily reliant on fossil fuels like coal. In such regions, there is little reduction potential from workflow shifting with and without interruptions. 

Given that our implementation of interrupted workflow shifting relies on workflow executions being divided into set execution windows, it may not find the `optimal' schedule for a workflow. However, it implements a form of interrupted workflow shifting that can outperform entire workflow shifting, or achieve savings in a shorter time -- highlighting its potential. 

\paragraph{Marginal CI}
When using the marginal signal, our results from interrupted workflow shifting were much less consistent, given that the signal exhibits less regular patterns in CI fluctuation, leading to shorter windows of low-carbon energy. 
In Germany, which typically has fluctuating CI (see Fig.~\ref{fig:carbon-intensity-fluctuation}), we see little reduction in the footprint of our workflows throughout the year. The most potential is observed in May, with a potential reduction of around 45\%, due to the presence of low-carbon intensity windows. 

\begin{figure}[h]
\centerline{\includegraphics[width=\columnwidth]{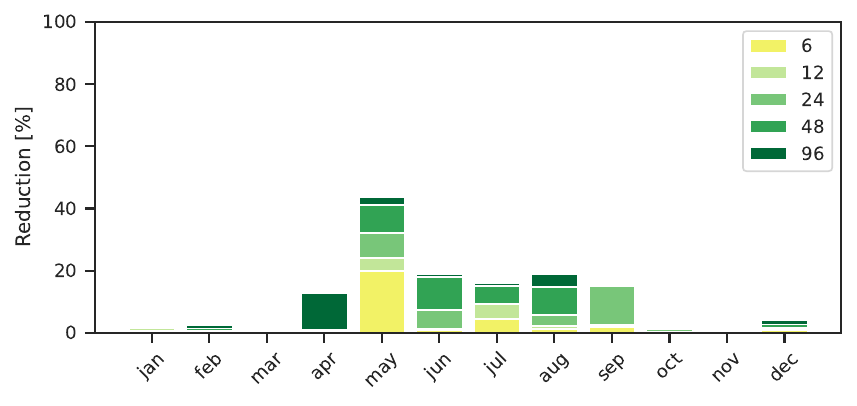}}
\caption{Footprint Reduction using Interrupted Shifting in Germany over windows of 6--96h.}
\label{fig:intshift-de-marg}
\end{figure}

\begin{figure}[h]
\centerline{\includegraphics[width=\columnwidth]{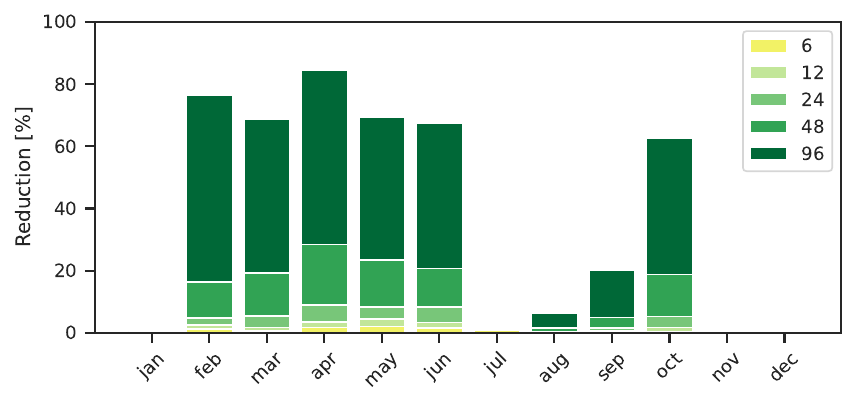}}
\caption{Footprint Reduction using Interrupted Shifting in California over windows of 6--96h.}
\label{fig:intshift-ca-marg}
\end{figure}

This is different for regions like California (see Fig.~\ref{fig:intshift-ca-marg}), which produces a significant amount of energy from solar. When using the marginal signal, we see potential for the footprint to be significantly reduced in February--June and October of 2024, with reductions of more than 70\% in the 96h window. 
However, in July--September, we note fewer low-carbon windows, potentially caused by the combination of increased energy usage and lower energy curtailment, resulting in a higher CI and less footprint reduction potential.

\paragraph{Summary}
Using marginal CI highlights the potential of leveraging a signal that could indicate energy curtailment or low grid demand, enabling scientists to execute workflows with essentially zero operational carbon emissions.

\paragraph{Storage Overhead from Interrupting Execution}
When we interrupt the workflow execution to wait for low-carbon energy to become available again, intermediate results must be stored on disk. Storing additional data on the disk may cause additional energy consumption and carbon emissions. 
We model the ``pessimistic'' data required by the workflows we study by summing all data read in and written out by each workflow task (a figure which includes files read in multiple times throughout execution, and files that would be discarded at different stages of execution). 
We found that the workflow data would fit on a single commercial storage disk utilized by a data center, with a capacity of 10TB+. A hard disk drive (HDD) would have a wattage of 7W, and a solid state drive (SSD) would have a wattage of 11W~\cite{shehabi20242024}. The results are shown in Table~\ref{table:int-shift-pause-overhead}.

We observed that as the length of the window explored increased, the time spent paused increased, leading to additional energy consumption. These models overestimate the energy, considering an entire hard disk -- in reality, these disks will typically be shared by workloads. Furthermore, most workflows we study required far less disk storage space, consuming less energy.

\begin{table}[htb]
\caption{Storage Overhead and Embodied Emissions (for a fully utilized enterprise SSD) from pausing scientific workflow execution.}
\label{table:int-shift-pause-overhead}
\centering
\rowcolors{2}{}{Gray}
\resizebox{\columnwidth}{!}{
\begin{tabular}{l|ccccc}
\toprule
Window (h) & 6 & 12 & 24 & 48 & 96 \\
\midrule
Mean Pause Duration (h) & 2 & 4 & 5 & 11 & 25 \\
HDD Energy (kWh) & 0.012 &  0.025 & 0.037 & 0.079 & 0.177 \\
SSD Energy (kWh) & 0.019 & 0.039 & 0.058 & 0.125 & 0.279 \\
Embodied Emissions ($gCO_2e$) & 2.86 & 5.72 & 7.15 & 15.73 & 35.75 \\
\bottomrule
\end{tabular}
}
\end{table}

\paragraph{Impact on Embodied Emissions}
When we perform interrupted workflow shifting, we introduce time overhead by pausing and resuming execution. This is because tasks that do not fully complete before pausing will need to be restarted when the workflow resumes. 
We observe a resultant increase in embodied carbon emissions of 0.03$g$. However, the mean reduction in average carbon emissions from applying interrupted shifting is 1.61$kg$. Therefore, the increase in embodied emissions from the CPU has a negligible impact on the overall carbon footprint. Furthermore, we expect an additional share of embodied carbon coming from the use of storage media to store intermediate data during the paused periods. 
Using the time spent paused as a fraction of the drive's expected lifetime, we consider an SSD that would be used in commercial clusters with a capacity of $\approx$15TB, namely the Micron 9300 Pro, which has an LCA footprint of $50.0 kgCO_2e$~\cite{Boavizta2025}. Assuming we use the entire drive, and the expected lifetime is 4 years, a workflow paused for 11 hours would cause 15.7$gCO2e$. The MAG workflow had the largest data write requirement, but would use less than 50\% of the drive. The baseline embodied emissions were 213.1$gCO2e$, so using 50\% of the drive and, therefore, adding 7.9$gCO2e$ would minimally increase the embodied emissions by 3.7\%, which is considerably less than the savings in operational emissions. Moreover, other workflows studied would use smaller shares of such a drive. 

\section{Potential of Resource Scaling}
\label{sec:resource-scaling-potential}
We explored the potential of carbon-aware scaling across two dimensions:
\begin{enumerate*}[label=(\roman*)]
    \item \textbf{Resource selection}, the impact of using different devices for individual workflow tasks; and 
    \item \textbf{Frequency scaling}, the impact of using different processor governors for individual tasks and entire workflows.
\end{enumerate*}
We also include an example of adjusting the cluster size for the execution of an entire workflow. 

We study the following workflow tasks: \textbf{bowtie2\_build}, \textbf{fastp}, \textbf{fastqc} and \textbf{trimgalore}. All four of these tasks are from the twenty most used bioinformatics tasks from Nextflow's community-curated nf-core library~\cite{ewels2020nf}. 

\subsection{Adjusting Compute Resources Used} 
In our first experiment, we explored the impact of choosing different nodes to execute the individual workflow tasks. We executed each task three times on each resource: three GCP nodes, three olympus nodes, elysium, camelot and sherwood (see Table~\ref{table:compute-resources}). 

To ensure our results were comparable, we adjusted the start time consistently for each task, where bowtie2\_build started at 09:00, fastp at 11:00, fastqc at 13:00 and trimgalore at 15:00. Each task was adjusted for the 'median' day of each month of the year (the day which fell in the middle of the month), in each of our selected regions. Task runtime, energy consumption, and carbon emissions from the mean of three executions on each node are shown in Table~\ref{table:task-resource-assign}. 

\begin{table*}[h]
\centering
\setlength{\tabcolsep}{3pt} %
\begin{tabular}{l|l|rrr|rrrrrr}
\toprule
Task & Dimension & \multicolumn{9}{c}{Machines} \\
& & gcp-c2 & gcp-n2 & gcp-n1 & olympus-1 & olympus-2 & olympus-3 & elysium & camelot & sherwood \\
\midrule
& runtime & 0.16 & \textbf{0.18} & 0.25 & 0.25 & 0.25 & 0.25 & 0.15 & 0.16 & \textbf{0.14} \\
bowtie2 & energy & 0.005 & 0.003 & \textbf{0.002} & 0.022 & 0.022 & 0.024 & 0.029 & 0.017 & \textbf{0.004} \\ 
\_build & avg. emis & 1.81 & 1.21 & \textbf{0.68} & 8.09 & 7.92 & 8.62 & 10.46 & 6.21 & \textbf{1.59} \\ 
& marg. emis & 1.45 & 1.09 & \textbf{0.74} & 8.82 & 8.66 & 9.47 & 7.40 & 4.90 & \textbf{1.03} \\ 
& emb. emis & \textbf{0.09} & 0.10 & 0.13 & 0.13 & 0.13 & 0.14 & 0.19 & 0.09 & \textbf{0.06} \\
\hline
\multirow{4}{*}{fastp} & runtime & \textbf{0.06} & 0.08 & 0.15 & 0.14 & 0.16 & 0.17 & 0.11 & \textbf{0.05} & \textbf{0.05} \\
& energy & 0.002 & 0.002 & \textbf{0.001} & 0.013 & 0.014 & 0.017 & 0.022 & 0.005 & \textbf{0.002} \\
& avg emis & 0.79 & 0.63 & \textbf{0.52} & 4.79 & 5.23 & 6.22 & 8.12 & 1.97 & \textbf{0.63} \\
& marg emis & 1.29 & 1.02 & \textbf{0.38} & 3.32 & 4.17 & 5.45 & 3.50 & 3.21 & \textbf{1.02} \\
& emb. emis & \textbf{0.03} & 0.04 & 0.08 & 0.08 & 0.09 & 0.09 & 0.15 & 0.03 & \textbf{0.02} \\
\hline
\multirow{4}{*}{fastqc} & runtime & \textbf{0.18} & 0.21 & 0.26 & 0.26 & 0.26 & 0.26 & \textbf{0.13} & 0.16 & 0.16 \\
& energy & 0.006 & 0.004 & \textbf{0.002} & 0.023 & 0.023 & 0.025 & 0.025 & 0.017 & \textbf{0.005} \\
& avg emis & 2.07 & 1.38 & \textbf{0.72} & 8.59 & 8.42 & 9.07 & 9.23 & 6.46 & \textbf{1.84} \\
& marg emis & 1.75 & 1.29 & \textbf{0.77} & 9.12 & 8.97 & 9.66 & 4.92 & 4.95 & \textbf{1.35} \\
& emb. emis & \textbf{0.10} & 0.11 & 0.14 & 0.14 & 0.14 & 0.14 & 0.17 & 0.09 & \textbf{0.07} \\
\hline
\multirow{4}{*}{trimgalore} & runtime & \textbf{1.07} & 1.28 & 1.93 & 1.50 & 1.49 & 1.56 & \textbf{0.86} & 1.05 & 1.0 \\
& energy & \textbf{0.002} & 0.006 & 0.009 & 0.046 & 0.044 & 0.053 & 0.170 & \textbf{0.005} & 0.021 \\
& avg emis & \textbf{0.89} & 2.00 & 3.23 & 16.44 & 15.63 & 19.06 & 61.38 & \textbf{1.80} & 7.66 \\
& marg emis & 20.30 & 14.28 & \textbf{10.68} & 77.69 & 75.08 & 84.53 & 95.97 & 62.47 & \textbf{17.7} \\
& emb. emis & \textbf{0.58} & 0.69 & 1.04 & 0.81 & 0.81 & 0.84 & 1.14 & 0.57 & \textbf{0.29} \\
\bottomrule
\end{tabular}
\caption{Resource Assignment results: each task's runtime is reported in hours, its energy in $kWh$, and its average (avg.), marginal (marg.), and embodied (emb.) emissions (emis), in $gCO_2e$, using CI data for Great Britain. Values in bold represent the minimum runtime, energy and emissions.}
\label{table:task-resource-assign}
\end{table*}

Across all nodes, the bowtie2\_build task took between 9--15m to run. The execution on elysium had the shortest runtime, but consumed the most energy, 0.029$kWh$. This was almost two times the energy consumed by camelot, which took longer but only consumed 0.017$kWh$. 
The choice of node significantly impacted the runtime of these tasks, with the trimgalore task taking between 52m--1h56m across all nodes. 

We note that the GCP machines show comparatively low energy consumption, and separate these results in the table. It is more challenging to accurately estimate energy consumption in the cloud, given that we could not perform power measurements on the nodes to fit power models. Therefore, we were reliant on average energy coefficients used by CCF~\textsuperscript{\ref{ccf-footnote}}, and believe that there is greater potential for discrepancies here. 

While the results from the experiment show that the sherwood and gcp-n1 nodes would allow for carbon emissions to be minimised (from each set of devices), changing the device that a task runs on can significantly impact the runtime, energy consumption and carbon emissions. It is important to consider workflow constraints such as scientists' deadlines or subsequent workflow tasks being dependent on the results produced by tasks to choose particular devices during workflow execution. Additionally, it is possible that task executions could be aligned with low-carbon intervals by selecting appropriate devices. 

\paragraph{Impact on Embodied Emissions}
Table~\ref{table:task-resource-assign} shows the estimated embodied carbon emissions of each task on each machine. Two factors impact the embodied emissions: the LCA emissions associated with each machine, and the runtime of the task on each machine. 
For the individual compute nodes, we found that sherwood minimised the embodied carbon emissions for all tasks, given that the node had the lowest LCA emissions and the lowest task runtimes. 
For the GCP nodes, gcp-c2 had the lowest emissions, given that it had the lowest task runtimes. 

\subsection{Adjusting Processor Governor Settings}  
Next, we explored the impact of frequency scaling, by running individual workflow tasks on nodes where the processor governor was changed. We focused on processor governor settings, as scientists might not have the permissions or expertise required to choose a specific frequency, and nodes may operate using pre-selected governors.

A processor's governor is a component that is used to manage the CPU clock speed in response to changes in system load. We focus on two Intel governors: \textbf{performance}, which forces the CPU to always run at the highest possible frequency; \textbf{powersave}, which forces the CPU to run at the lowest possible frequency. 
For each selected resource, we took separate power measurements as a basis for estimating energy consumption for each governor setting. 

We executed the tasks on three olympus nodes, elysium, camelot and sherwood, as presented in Table~\ref{table:compute-resources}. 
For all executions of the selected tasks, we adjusted the start times to the same as that of the resource assignment experiment, again using the median day in each month. In our discussion we use the estimated carbon emissions using CI data for Great Britain. 

In Table~\ref{table:task-freq-scale-nodes}, we compare the nodes elysium and camelot. The elysium node is the most powerful and newest node that we studied, while camelot is several years older. 
We observed that for elysium, changing the governor from performance to powersave had little impact on the runtime of each task -- with the powersave governor consuming slightly less energy, and producing slightly less carbon emissions. 
In contrast, we observe a far greater difference between governor settings for camelot. The runtime of each task is around four times longer when using the powersave governor. Given the increase in task runtime, using the powersave governor consumes around twice the energy of the performance governor. On this node, we might therefore generally prefer to use the performance governor to reduce our runtime, energy consumption, and carbon emissions. 
Furthermore, when comparing these nodes, using camelot with the performance governor would offer the minimum energy consumption and carbon emissions, despite not always having the shortest runtime. 

\begin{table}[h]
\centering
\setlength{\tabcolsep}{4pt} %
\resizebox{\columnwidth}{!}{
\begin{tabular}{l|l|rr|rr}
\toprule
& & \multicolumn{2}{c}{Node elysium} & \multicolumn{2}{c}{Node camelot} \\
Task & Dimension & \multicolumn{2}{c}{Governor} & \multicolumn{2}{c}{Governor} \\
& & perf. & save. & perf. & save. \\
\midrule
& runtime & \textbf{0.15} & \textbf{0.15} & 0.16 & 0.64 \\
bowtie2 & energy & 0.029 & 0.027 & \textbf{0.017} & 0.037 \\
\_build & avg. emis & 3.25 & 2.65 & \textbf{1.93} & 3.53 \\
& marg. emis  & 5.07 & 11.37 & \textbf{3.36} & 14.87 \\
& emb. emis & 0.16 & 0.16 & \textbf{0.07} & 0.28 \\
\hline
\multirow{4}{*}{fastp} & runtime & 0.11 & 0.11 & \textbf{0.05} & 0.20 \\
& energy & 0.022 & 0.022 & \textbf{0.005} & 0.011 \\
& avg. emis & 2.21 & 2.32 & \textbf{0.54} & 1.03 \\
& marg. emis & 2.36 & 9.03 & \textbf{2.19} & 4.61 \\
& emb. emis & 0.12 & 0.12 & \textbf{0.02} & 0.08 \\
\hline
\multirow{4}{*}{fastqc} & runtime & \textbf{0.13} & \textbf{0.13} & 0.16 & 0.63 \\
& energy & 0.025 & 0.024 & \textbf{0.017} & 0.036 \\
& avg. emis  & 2.42 & 2.86 & \textbf{1.69} & 4.84 \\
& marg. emis  & 3.52 & 10.35 & \textbf{3.54} & 16.11 \\
& emb. emis & 0.13 & 0.14 & \textbf{0.07} & 0.27 \\
\hline
\multirow{4}{*}{trimgalore} & runtime  & \textbf{0.86} & \textbf{0.86} & 1.05 & 4.27 \\
& energy & 0.170 & 0.160 & \textbf{0.005} & 0.246 \\
& avg. emis & 17.52 & 19.31 & \textbf{0.55} & 26.33 \\
& marg. emis & 66.27 & 67.49 & \textbf{43.31} & 99.16 \\
& emb. emis & 0.92 & 0.92 & \textbf{0.02} & 1.42 \\
\bottomrule
\end{tabular}
}
\caption{Frequency Scaling for elysium and camelot using the performance (perf.) and powersave (save.) governors: each task's runtime is reported in hours, its energy in $kWh$, its average (avg.), marginal (marg.), and embodied (emb.) emissions in $gCO_2e$ using CI data from Great Britain. Values in bold represent the minimum runtime, energy and emissions.}
\label{table:task-freq-scale-nodes}
\end{table}

However, the task-level experiment did not consider the impact of frequency scaling on an entire workflow. 
So, our next experiment considered the execution of full workflows, Chip-Seq and RNA-Seq, exploring the impact of the same governors on overall energy consumption and carbon emissions. The results are shown in Table~\ref{table:workflow-freq-scale-avg}. We used the average CI for all selected regions to consider the impact across the world. 

Both workflows were executed on a cluster of eight camelot nodes. 
The Chip-Seq workflow took around 3h18m to execute using the performance governor, consuming 3.76$kWh$ of energy, and around 8h30m to execute using the powersave governor, consuming 5.37$kWh$ of energy. 
Meanwhile, the RNA-Seq workflow took around 2h24m to execute using the performance governor, consuming 2.37$kWh$ of energy, and around 7h30m to execute on the powersave governor, consuming 3.34$kWh$ of energy. 
We observe that using the powersave governor results in runtimes of around 2.5--3x the duration of the performance governor when these workflows are executed on this cluster. However, the energy consumption is only around 1.4x higher using the powersave governor. While consuming more overall energy leads to greater carbon emissions using the average CI signal in most of our regions, we notice that the Chip-Seq workflow running in California and Texas, we would produce less emissions by using the powersave governor. Therefore, the processor's governor setting could be adjusted to control the energy consumption over time, to align with low carbon energy available in a region and reduce overall carbon emissions. 

\begin{table*}[]
\centering
\setlength{\tabcolsep}{4pt} %
\begin{tabular}{llr|rrrrrrr}
\toprule
& & & \multicolumn{7}{c}{Regions} \\
& & & Great & & & & South & & New South \\
Workflow & Governor & Energy & Britain & Germany & California & Texas & Africa & Tokyo & Wales \\
\midrule
\multirow{2}{*}{Chip-Seq} & perf. & 3.76 & \textbf{190.09} & \textbf{846.69} & 751.91 & 1,844.90 & \textbf{2,443.08} & \textbf{1,571.35} & \textbf{2,164.10} \\
& save. & 5.37 & 538.06 & 1,084.80 & \textbf{722.09} & \textbf{629.96} & 3,492.61 & 2,477.60 & 2,873.44 \\
\hline
\multirow{2}{*}{RNA-Seq} & perf. & 2.37 & \textbf{92.10} & \textbf{400.80} & \textbf{185.74} & \textbf{1,108.29} & \textbf{1,545.18} & \textbf{1,148.62} & \textbf{1,482.18} \\ 
& save. & 3.34 & 203.62 & 650.07 & 338.57 & 2,443.08 & 2,202.84 & 1,431.15 & 1,820.24 \\ 
\bottomrule
\end{tabular}
\caption{Frequency Scaling for Chip-Seq and RNA-Seq workflows using the performance (perf.) and powersave (save.) governors in all regions using average. Values in bold represent the minimum emissions for each workflow in each region.}
\label{table:workflow-freq-scale-avg}
\end{table*}

\paragraph{Impact on Embodied Emissions}
Table~\ref{table:task-freq-scale-nodes} shows the estimated embodied carbon emissions of each task running with the performance and powersave governors on each machine. We observe that changing the governor impacts the runtime of each task, impacting the embodied emissions. 
Using camelot with the performance governor resulted in the lowest embodied carbon emissions for all tasks due to the machine's lower LCA emissions ($21kgCO_2e$ compared to $46.7kgCO_2e$ for the elysium machine), despite having longer task runtimes. For entire workflow execution, we observed that using the performance governor resulted in lower embodied carbon emissions than the powersave governor, as it reduced the workflow's runtime by 74\%%
, which is reflected in the reduction in embodied emissions. 

\subsection{Adjusting Cluster Size for Entire Workflow Execution}
We additionally used workflow traces to explore the execution of the Chip-Seq workflow on a cluster formed of 2, 4 and 8 atlantis nodes using CI data for Germany during October to December 2023. 

\begin{table}[htb]
\caption{Carbon Emissions for Chip-Seq on different cluster sizes. Values in bold represent the minimum runtime, energy and emissions.}
\label{table:cluster-scaling}
\centering
\rowcolors{3}{Gray}{}
\resizebox{\columnwidth}{!}{
\begin{tabular}{lrrrrr}
\toprule
\# & Runtime & Energy & Avg. emis & Marg. emis & Emb. emis \\
Nodes & (h)  & (kWh) & (gCO2e) & (gCO2e) & (gCO2e) \\
\midrule
2     & 11.84 & \textbf{6.78} & 2,341.01 & 5,054.35 & \textbf{15.65} \\
4     & 5.97 & 6.82 & 1,913.76 & 5,132.89 & 15.78 \\
8     & \textbf{3.13}& 6.94 & \textbf{1,376.98} & \textbf{4,425.68} & 16.55 \\
\bottomrule
\end{tabular}
}
\end{table}

As shown in Table~\ref{table:cluster-scaling}, the runtime reduced as the number of nodes increased. The workflow consumed similar amounts of energy, marginally rising as the number of nodes increased. These decreases in runtime lead to a reduction in the carbon footprint due to the given CI data. However, further reductions could be made by aligning shorter executions optimally with low-carbon windows of electricity. 

\paragraph{Impact on Embodied Emissions}
As we use more nodes, we must consider the embodied emissions of additional nodes, but at the same time, the workflow runtime decreases significantly, leading to a lower share of LCA emissions. Consequently, we observe that the embodied emissions increase slightly in Table~\ref{table:cluster-scaling}. 

\section{Discussion}
In this section, we discuss the key takeaways from our experiments and threats to the validity of our evaluation. 

\subsection{Key Takeaways}

\paragraph{Carbon-Aware Workflow Shifting}
In Section~\ref{Subsec:DelayTolerance}, we shifted the execution of entire workflows within a flexibility window to assess the potential reduction in carbon emissions. As the length of this flexibility window was increased, the potential reduction in carbon emissions also increased. This demonstrates that the delay tolerance of many scientific workflows can be leveraged for low-carbon execution. While temporal shifting achieved carbon emission reductions in most of our selected regions, some regions, such as South Africa, exhibited minimal reductions due to their lower renewable energy generation and, accordingly, fewer fluctuations in the energy mix compared to regions like Great Britain. 

\begin{finding}
In regions with a significant presence of renewable energy sources and fluctuating CI, temporal shifting of entire workflows could result in carbon footprint reductions of over 80\% using average CI. 
\end{finding}

\paragraph{Interruptible Workload Shifting}
In Section~\ref{Subsec:Interruptibility}, we went beyond shifting entire workflows by dividing execution into execution windows that could be mapped to multiple lowest-carbon CI intervals. 
These experiments found that interrupted workflow shifting could achieve greater savings in carbon emissions in a shorter time than entire workflow shifting. 
In particular, California showed the potential for reductions of 30--70\% in a flexibility window of 6--24h, and up to 80\% in a window of 96h, throughout the year. This significantly improves over the reductions possible for California when shifting entire workflows. 

Moreover, our experiments with Marginal CI highlighted that the choice of signal used when shifting is important, and impacts potential reductions. The marginal signal could indicate periods of time where energy is curtailed, or the grid has low demand, to execute workflows with little-to-no operational emissions. 

\begin{finding}
Shifting workflows with interruptions can amplify savings, e.g., 30--70\% in a 24h window, improving from reductions of $<$20\% using entire workflow shifting in the same window for all workflows within regions that have a significant presence of renewable energy, using average CI. 
\end{finding}

\paragraph{Carbon-Aware Resource Scaling}
In Section~\ref{sec:resource-scaling-potential}, we demonstrated how choosing different nodes impacted the runtime, energy consumption and carbon emissions of four tasks. We found that the choice of device significantly changed these properties of tasks. If we wanted to execute a workflow in a carbon-aware manner, we could choose to run tasks on specific devices to closely align with low-carbon windows.  
Next, we explored the impact of frequency scaling, comparing the powersave and performance governors at an individual task and entire workflow granularity. Here, we saw that each governor offered different benefits. Powersave tended to take longer but consumed less energy over the same period of time, while performance led to more work being completed in a shorter time, but consumed more energy to achieve this goal. Alternating between these governors could lead to a workflow's energy consumption being adjusted depending on the region's CI. 
We also showed that adjusting the cluster size for the Chip-Seq workflow could significantly impact runtime to highlight the potential for carbon-aware resource scaling for entire workflows. For example, more computing resources could be allocated to low-carbon windows to reduce the workflow footprint. 
Such methods could be combined with carbon-aware interruptible shifting to divide workloads into groups of tasks that maintain workflow data dependencies, while making the best use of forecasted low-carbon windows.  

\begin{finding}
Resource scaling can shape runtimes of workflows and their tasks to fit upcoming low-carbon energy availability, e.g., if Chip-Seq used the performance governor instead of powersave, it could reduce carbon emissions by 67\% on the same compute cluster. %
\end{finding}

\subsection{Threats to Validity}
\label{Sec:threats}

\paragraph{Power Estimation vs. Measurement}
To estimate workflow/task carbon footprint, we used readily available Nextflow trace files to avoid unnecessary emissions. This necessitated the use of a linear power model, which is inherently less accurate than directly monitoring via hardware or software power meters~\cite{10171575,west2025ichnoscarbonfootprintestimator}. However, the consistent use of this methodology across both baseline and experimental scenarios ensures that the relative reductions remain representative. Moreover, widely used footprint assessment methodologies (e.g., CCF\textsuperscript{\ref{ccf-footnote}} and Green Algorithms\textsuperscript{\ref{ga-footnote}}) similarly rely on linear models. 

\paragraph{Assumption of Perfect Knowledge}
We estimated the footprint for all possible time-shifting scenarios with and without interruptions. This approach assumed we had perfect knowledge of workflow runtimes and error-free CI forecasts. These assumptions do not necessarily reflect real-world conditions. 
However, many schedulers are reliant on such signals~\cite{Topcuoglu_2002_HEFT,DURILLO2014221,8400402}, and existing methods used to predict the runtime and energy consumption of workflow tasks can have a low error, with \cite{hilmanTaskRuntimePrediction2018,huangCloudProphetMachineLearningBased2024,BADER2024171} suggesting 3--30\%.
Meanwhile, CI forecast services like ElectricityMaps and WattTime are widely used and report a 10--15\%\footnote{\url{https://www.electricitymaps.com/technology}} and 1--9\%\footnote{\url{https://watttime.org/data-science/methodology-validation/}} errors, respectively.
We conducted a sensitivity analysis to explore the impact of inaccurate CI forecasting with a 5, 10, and 15\% errors, and inaccurate task runtime predictions with a 10, 15, and 25\% errors. In both experiments, we observed that making predictions with inaccurate forecasts made minimal differences to the potential carbon footprint reduction. 

\paragraph{Assumption of Infinite Compute Resources}
Throughout our experiments, we assumed unconstrained resource availability, allowing for maximum exploitation of low-carbon energy windows. While this does not reflect all real-world operational settings, we identify two scenarios where access to effectively `infinite' compute resources is feasible. 
The first scenario stems from our own experience of using local CPU compute clusters at four universities that were considerably underutilized, and therefore had the capacity to support all considered techniques. We believe this to not be an uncommon situation for research organizations. 
The second scenario has scientists utilize public cloud resources, which offer on-demand access to massive resource pools. With public cloud providers consistently offering surplus resources with high availability through spot markets~\cite{9975369}, resources are likely to be available on-demand at all times. Beyond that, empirical studies indicate that cloud compute resources, such as CPU and memory, are frequently underutilized~\cite{van2022scalable,guo2019limits}. 
With such capacities, there is potential for the application of carbon-aware computing techniques. 
However, we acknowledge that in settings of capacity constrains, this potential cannot be realized.

\paragraph{Assumption of Efficient Interruption}
We assume that intermediate data generated by workflow tasks would be stored on a networked or distributed file system, allowing for execution to be paused and resumed with minimal overhead. 
Our work focused on Nextflow SWMS, which exchanges intermediate data through the file system. However, the assumption also holds for SWMSs like Pegasus, Snakemake, Makeflow, and Argo which are also capable of running on a networked or distributed file system, and allow for workflow execution to be paused and resumed with limited overhead, beyond task alignment.

\paragraph{Storage Overhead from Interrupting Execution}
Interrupting workflow execution requires intermediate task results to be stored on disk, which incurs additional cost. 
To quantify this, we modeled the energy consumed by a HDD and SSD, utilizing the entire drive for the average time our workflows were paused, in Table~\ref{table:int-shift-pause-overhead}.  
Our analysis indicates that while the energy consumed increases with the duration of the flexibility window, the overall impact remains minimal. Even for the most data-intensive workflows like MAG, which features the highest data write requirement, the original energy consumed was 21.12$kWh$; hence, temporarily storing intermediate data on an SSD while the execution is paused would increase the energy consumption by only 1.4\%, and for RNA-Seq: 5.43$kWh$ and 5.3\% under pessimistic assumptions.
Moreover, the commercial disk modeled would typically have a capacity of 10TB or more, and would be shared between users, leading to an even smaller share of energy consumption attributed to the workflow. 
In terms of embodied carbon, we used the time spent paused as a fraction of the drive's lifetime, and the share of drive capacity. For instance, pausing for 11 hours would increase MAG's embodied carbon by 3.7\% and RNA-Seq's by 1.1\%. 
While the storage-related emission overheads are negligible within the scope of our study, the overhead should be considered when optimizing data intensive workflows.

\paragraph{Coverage of Renewable Energy Signals}
In our experiments, we estimated the carbon footprint using hourly average and marginal CI data, as most granular historical data were only available at an hourly level. This aligns with established research in carbon-aware computing %
(Section~\ref{sec:RelatedWork}). 
Future work should incorporate higher-resolution data as more granular CI forecasting becomes available. %
In addition, while other metrics for renewable energy accounting exist (e.g., market-based measures such as RECs and PPAs), we did not focus on these as they do not necessarily correlate with the physical availability of low-carbon energy at a certain time and location.

\paragraph{Workflow System Representativeness}
While this study only focuses on a single SWMS (Nextflow), the highlighted properties (delay tolerance, interruptibility, scalability, and heterogeneity) are common across many SWMSs. Therefore, the observed reductions in emissions from applying carbon-aware methods are likely generalizable to other SWMSs. 

\paragraph{Realistic System-Level Restart Overheads}
We conducted a short experiment to measure the system-level cost of interrupting a Nextflow workflow execution using Kubernetes by running the RNA-Seq workflow with a test dataset, so that the interruption overhead stems mainly from the job submission, task scheduling, and container runtime restarting. We measured the execution time when run continuously, which was 486s, and the the execution time when it was interrupted and resumed, which was 582s. Here, the runtime increased by 96s. For a short workflow, this is a significant increase in makespan (20\%), but for workflows that run for several hours, the system-level overhead is far less significant, e.g. +1\% for a 3h execution.

\paragraph{Scale of Footprint Reduction}
Our evaluation demonstrates the potential for substantial emission savings, e.g., kilograms of carbon emissions for medium-scale workflows that run for up to 12h. While these workflows may have a smaller individual footprint than those running for thousands of core hours (e.g.,~\cite{deelmanPegasusCloudScience2016}), they are frequently executed by scientists. 
We expect these savings to scale proportionally to larger computational tasks, especially for larger infrastructures and more expansive flexibility windows. 

\paragraph{Absolute Carbon Emission Savings and Trade-offs}
Our evaluation showed significant potential for carbon emissions to be reduced, but the trade-offs of each mechanism and absolute emission savings must be considered.

When applying temporal shifting without interruptions, using average CI, we could reduce emissions by 45+\% for the region of Great Britain. For a workflow with a footprint of~$\approx$4$kgCO2e$, a 45\% reduction is 1.8$kgCO2e$. For others with a footprint of~$\approx$0.2$kgCO2e$, the same reduction is 0.1$kgCO2e$. The absolute reduction depends on the workflow executed and the resources utilized. 
With interruptions, 40--65\% savings could be achieved for the region of California within a 12h window.
When a workflow is delayed for a day or over the work week, the energy consumed does not change. With interruptions, the energy consumed can increase from pausing and resuming execution, and has a storage overhead, but greater emission savings can be made in a shorter time, e.g. 12h instead of 96h.

When applying resource scaling, we found that executing workflows on devices with either the performance or powersave governor changed the energy consumed and emissions:
Running RNA-Seq with the performance governor reduced footprint by 0.3$kgCO2e$ in the Tokyo region, and by 1.3$kgCO2e$ in Texas. The absolute reduction potential depends on the region where the workflow was executed -- different carbon-aware mechanisms may be more effective in certain regions.
Using the performance governor reduced the runtime, but increased the rate of energy consumed over time.

With each mechanism, absolute carbon emission savings must be balanced with trade-offs like delays, increased energy consumption, and storage requirements.

\paragraph{Embodied Carbon Proportional Runtime Attribution}
Time-shifting and workflow interruption optimizations to operational emissions may incentivise companies to invest in additional compute and storage to take advantage of the lower CI windows. This comes with additional embodied emissions. It may also result in hardware underutilization during higher CI windows, which would lead to an increase in embodied emissions not attributed to any workload. Not reflecting such rebound effects is a known limitation of the proportional runtime attribution methodology and needs to be addressed. It is, however, outside of the scope of this work.

\section{Related Work}
\label{sec:RelatedWork}

\paragraph{Temporal Shifting}
Various works explored scheduling or interrupting applications to synchronize consumption with periods of low CI. Through simulations, Wiesner et al.~\cite{wiesnerLetWaitAwhile2021c} demonstrated the efficacy of delaying or interrupting flexible workloads, such as ML training. %
Industrial implementations, such as Google's Carbon-Intelligent Compute Management~\cite{9770383}, use average CI forecasts to set capacity limits for data centers, thereby deferring flexible workloads.
Similarly, Lin et al.~\cite{lin2023adapting} developed capacity plans for hyper-scale data centers using day-ahead forecasting to mitigate grid instability.
Piontek et al.~\cite{Piontek2024} used historical CI data to align Kubernetes jobs with predicted low-carbon energy.
These approaches reduce emissions through temporal shifting, but focus on individual workloads, rather than workflows of inter-dependent tasks.

Other studies aim to leverage surplus renewable energy directly. Zheng et al.~\cite{zheng2020mitigating} investigated load migration between data centers to utilize energy that would otherwise be curtailed. Cucumber~\cite{10.1007/978-3-031-12597-3_14} introduced a configurable admission control policy specifically for delay-tolerant workloads in edge data centers with on-site renewable sources, while FedZero~\cite{10.1145/3632775.3639589} restricts Federated Learning tasks to periods of excess renewable energy and spare compute capacity. Crucially, none of these approaches specifically address scientific workflows.
Lechowicz et al.~\cite{PCAPS2025} propose a scheduler to align tasks with precedence with the availability of low-carbon energy.
Bostandoost et al. \cite{bostandoost2025quantifyingcarbonreductiondag} model potential carbon savings when applying temporal shifting without interruptions, incorporating trade-offs between emissions, energy, and makespan. They compute upper bounds assuming perfect knowledge of task dependencies, runtimes, and CI. They, too, note significant potential for emissions to be reduced. 
Schweisgut et al.~\cite{schweisgut2025carbon} proposed scheduling algorithms to align a pre-existing mapping and ordering of workflow tasks with low-carbon energy, though their focus remains on temporal shifting rather than the interplay of interrupted shifting and multi-device resource scaling.

\paragraph{Resource Scaling}
Carbon-aware resource scaling involves dynamically allocating more resources when CI is low, and reducing demand when it is higher. CarbonScaler~\cite{10.1145/3626788} performs horizontal scaling based on one-time offline profiling.
However, because it determines scaling factors from short execution windows, it may fail to capture the evolving demands of multi-stage workflows.
Conversely, Carbon Containers~\cite{10.1145/3620678.3624644} integrates vertical scaling, container migration, and temporal shifting to control and limit the emissions rate of containerized applications. While effective for isolated services, this method is not directly applicable to multi-stage applications like scientific pipelines.

\paragraph{Other Carbon-Aware Computing Techniques}
Broader frameworks have also emerged to prioritize sustainability across the compute stack. 
Carbon Explorer~\cite{10.1145/3575693.3575754} predicts carbon-optimal strategies for operating data centers by balancing renewable energy investment, energy storage, and carbon-aware shifting. 
Chien et al.~\cite{chien2023reducing} studied the impact of location-based shifting on the emissions generated by generative AI.
Ecovisor~\cite{10.1145/3575693.3575709} virtualizes the energy system to delegate renewable energy management to directly to the application.
Wen et al.~\cite{9298863} proposed increasing the usage of green energy when executing industrial workflows via location-based load shifting; however, their algorithm relies on assigning data centers static measures of energy mix `greenness' rather than the dynamic, time-varying nature of renewable energy generation. 
Lotaru~\cite{bader2022lotaru} is a method for predicting workflow runtime using microbenchmarks, and has been proven effective for simple carbon-aware time shifting~\cite{BADER2024171}.
Souza et al.~\cite{CASPER2024} create a system to provision resources and distribute web requests based on CI and availability in different regions to reduce carbon emissions. 
Rodrigues et al.~\cite{rodrigues2025carbonawaretemporaldatatransfer} present a scheduler that align and scale bulk data transfers with low-carbon energy.
Crucially, none of these techniques specifically exploits the characteristics of workflows for carbon-aware execution considering both temporal shifting and resource scaling.

In previous work~\cite{11044837}, we highlighted that scientific workflows are inherently delay tolerant, interruptible, scalable and heterogeneous; properties that increase the aptitude for carbon-aware execution of scientific workflows. We conduct a significantly more rigorous evaluation in this study, involving a larger number of workflows, varying tasks, and diverse geographical regions.

\section{Conclusion} 
In this paper, we have systematically explored the potential of carbon-aware execution for scientific workflows. 
By quantifying the environmental impact of seven real-world workflows, we established that baseline operations generate up to 17$kg$ of operational carbon emissions. 
Using these estimates as baselines, we first assessed carbon-aware time shifting, demonstrating footprint reductions of over 80\% using the average CI, and in some scenarios, completely using the marginal CI.
Our results indicate that interrupted shifting further amplifies savings, potentially outperforming full workflow shifting over the same flexibility windows. 

Beyond temporal adjustments, we evaluated resource assignment and frequency scaling, revealing that strategic device selection aligns execution with low-carbon windows. Furthermore, using different processor governors yielded a footprint reduction of 67\% under average CI.  
Based on these findings, we discussed how three intrinsic properties of scientific workflows -- namely delay tolerance, interruptibility, and scalability -- could be effectively leveraged for carbon-aware computing. 

Despite these promising results, our evaluation assumed perfect knowledge of task and workflow execution, flawless CI forecasts, and infinite resource availability. 
Consequently, our study does not fully account for practical implementation constraints. 
Future work will therefore explore the integration of real-world profiling techniques %
to measure and predict workflow performance, focusing on their runtime, energy consumption and expected carbon footprint on heterogeneous infrastructure. We would then explore how workflow profiles could be combined with CI forecasting and resource availability data to execute workflows in a carbon-aware manner, in practice. 

\section*{Acknowledgments}
This work was supported by the Engineering and Physical Sciences Research Council (EPSRC) under grant EP/UKRI154 (``Casper: Carbon-Aware Scalable Processing in Elastic Clusters'') and the German Research Council (DFG) as part of CRC 1404 (``FONDA: Foundations of Workflows for Large-Scale Scientific Data Analysis''). We also express gratitude to the providers of electricity grid data: NESO Open Data, and Electricity Maps for average carbon intensity; and WattTime for marginal operating emission rates.

\bibliographystyle{elsarticle-num}
\bibliography{refs}

\end{document}